\documentclass[a4paper]{article}

\usepackage{multicol}
\usepackage{amsthm}
\usepackage{bm}

\usepackage[dvipdfmx]{graphicx}
\usepackage{bmpsize}
\usepackage{amssymb}
\usepackage{amsmath}
\usepackage{color}

\definecolor{dgreen}{rgb}{0.0, 0.5, 0.0}

\begin{document}

\fontsize{14pt}{28pt}\selectfont

\begin{center}
\bf{Dynamical properties of max-plus equations \\
  obtained from tropically discretized Sel'kov model}
\end{center}\ \\
\fontsize{12pt}{12pt}\selectfont
\begin{center}
Shousuke Ohmori$^{*)}$ and Yoshihiro Yamazaki\\ 
\ \\
\it{Department of Physics, Waseda University, Shinjuku, Tokyo 169-8555, Japan}\\
\ \\
*corresponding author: 42261timemachine@ruri.waseda.jp\\
\end{center}
~~\\
\rm
\fontsize{11pt}{14pt}\selectfont\noindent

\baselineskip 30pt

{\bf Abstract}

Max-plus equations are derived from tropically discretized 
Sel'kov model via ultradiscretization. 
These max-plus equations possess common dynamical structures 
with the discretized model: 
Neimark-Sacker bifurcation and limit cycles. 
The limit cycles of the ultradiscrete max-plus equations 
have seven discrete states.
Furthermore, these max-plus equations exhibit excitability.
Relationship between the tropically discretized model 
and the derived max-plus equations is also discussed 
based on numerical results.
It is found that the derived max-plus equations correspond to 
a limiting case of the tropically discretized model.\\

\hrulefill
%
%
%



%
%

%
%
%

Ultradiscretization is known to be a method to 
derive max-plus equations from difference equations.
Actually for example, ultradiscretization has been applied 
to a discretized KdV equation.
As a result, a max-plus equation has been obtained, 
which is often referred as the box-and-ball system 
\cite{Tokihiro1996,Tsujimoto1998}.
The box-and-ball system (ultradiscrete system) 
exhibits soliton behaviors which are also confirmed 
in the original KdV equation (continuous system).
%
%
The important point is that ultradiscretization 
successfully retains and elucidates 
the essence of dynamical structures 
in the original nonlinear systems 
(soliton behaviors in this example), 
although the ultradiscrete system does not reproduce 
all dynamical properties of the original system.
%


Non-integrable dissipative systems 
and reaction-diffusion systems 
have also been target for application of ultradiscretization 
\cite{Nagatani1998,Murata2013,Ohmori2014,Matsuya2015,Murata2015,Gibo2015,Ohmori2016,Ohmori2020}.
Recently, we have applied ultradiscretization 
to bifurcation phenomena in one-dimensional dynamical systems\cite{Ohmori2020}.
Bifurcation phenomena have been hugely studied 
from viewpoint of continuous\cite{Guckenheimer,Strogatz,Nicolis} 
and discrete\cite{Robinson,Kuznetsov, Galor} dynamical systems.
In our previous study, focusing on tropically discretized 
one-dimensional normal forms 
of saddle-node, transcritical, and pitchfork bifurcations, 
their ultradiscrete max-plus equations were derived 
and their dynamical properties were investigated.
In particular, it is found that they possess 
``ultradiscrete bifurcations'', 
which are piecewise linear bifurcations 
for saddle-node, transcritical, and pitchfork types, respectively. 
%
%
These ultradiscrete bifurcations can be visually understood 
by piecewise linear graphs for the ultradiscrete 
max-plus equations.
%

%

%
In this letter, we focus on Neimark-Sacker bifurcation 
(N-S bifurcation hereafter) 
in two-dimensional dynamical systems 
by considering tropically discretized Sel'kov model 
as an example.
(N-S bifurcation is often referred as Hopf bifurcation for maps 
or discrete dynamical systems.)
From this discretized model, max-plus equations are derived 
by ultradiscretization. 
We show that the obtained max-plus equations exhibit 
N-S bifurcation, excitability, and existence of limit cycles.
Furthermore, relationship between the discretized model and 
the derived max-plus equations is numerically discussed.
%

Let us start with introducing 
the tropically discretized Sel'kov model for $(x_{n}, y_{n})$: 
\begin{eqnarray}{}
	x_{n+1} = \frac{x_n+\tau(ay_n+x^2_ny_n)}{1+\tau}, 
	 \label{eqn:3-2a}
\\
 y_{n+1} = \frac{y_n+\tau b}{1+\tau(a+x_n^2)}, 
	 \label{eqn:3-2b}
\end{eqnarray}
where $a$, $b$ and $\tau$ are positive parameters. 
The positive integer $n$ shows the iteration steps.
It is confirmed that in the limit of $\tau \rightarrow 0$ 
these equations correspond 
to the continuous Sel'kov model \cite{Strogatz,Selkov1968}: 
\begin{eqnarray}{}
  		\displaystyle\frac{d x}{d t}
    		& = & -x+ay+x^2y, 
  		\label{eqn:3-1a}
  \\
 	 \displaystyle\frac{d y}{d t}
    & = &  b-ay-x^2y, 
		\label{eqn:3-1b}
\end{eqnarray}
where $x_n=x(n\tau) \rightarrow x(t)$ 
and $y_n=y(n\tau) \rightarrow y(t)$ for $\tau \rightarrow 0$.
In eqs.(\ref{eqn:3-1a}) and (\ref{eqn:3-1b}), 
Hopf bifurcation occurs when $a$ and $b$ satisfy 
$\displaystyle b^2 = \frac{1}{2}(1-2a\pm \sqrt{1-8a}) 
\left( \equiv b_{\pm}^{2}(a) \right)$, 
and limit cycle solutions are obtained 
for $b_{-}(a) < b < b_{+}(a)$.

In general, the tropical discretization can avoid 
appearance of negative values in the equations; 
otherwise the negative values make ultradiscretization impossible \cite{Murata2013}.
Ultradiscretization for eqs.(\ref{eqn:3-2a})-(\ref{eqn:3-2b}) 
is performed in the following way.
The variable transformations, 
\begin{equation}
	\tau = e^{T},\;\; x_n = e^{X^{\prime}_n},\;\; 
	y_{n}= e^{Y^{\prime}_{n}},\;\; a  = e^{A},\;\; b =e^{B^{\prime}}, 
	\label{eqn:3-3}
\end{equation}
are applied to eqs.(\ref{eqn:3-2a})-(\ref{eqn:3-2b}), 
and the ultradiscretization 
	\begin{equation}
		\ln (e^{\alpha}+e^{\beta}+\cdot \cdot \cdot ) \rightarrow \max(\alpha, \beta,\dots)
		\label{eqn:0}
	\end{equation}
is carried out \cite{ultradiscrete}. 
Then, we obtain 
	\begin{eqnarray}
		X^{\prime}_{n+1} & = & \max(X^{\prime}_n, T+\max(A+Y^{\prime}_n,2X^{\prime}_n+Y^{\prime}_n))-\max (0,T),
		\label{eqn:3-4a} \\
		Y^{\prime}_{n+1} & = & \max(Y^{\prime}_n,T+B^{\prime})-\max(0,T+\max(A,2X^{\prime}_n)).
		\label{eqn:3-4b}
	\end{eqnarray} 
Additionally assuming as the condition of $T$ that 
\begin{equation}
	T \geq \max\{0,-A, Y^{\prime}_n-B^{\prime}, -(X^{\prime}_n+Y^{\prime}_n)\}
	\label{eqn:T-condi}
\end{equation}
for all $n$, eqs. (\ref{eqn:3-4a})-(\ref{eqn:3-4b}) become 
\begin{eqnarray}
	X^{\prime}_{n+1} & = & Y^{\prime}_n + \max(A,2X^{\prime}_n),
	\label{eqn:2-1a} \\
	Y^{\prime}_{n+1} & = & B^{\prime}-\max(A,2X^{\prime}_n).
	\label{eqn:2-1b}
\end{eqnarray} 
%
%
After the variable transformations 
\begin{equation}
	\displaystyle X^{\prime}_{n}-\frac{A}{2} = X_{n}, \;\;\;
	\displaystyle Y^{\prime}_{n}+\frac{A}{2} = Y_{n}, \;\;\;
	\mbox{and} \;\;\;
	\displaystyle B^{\prime}-\frac{A}{2} = B, 
	\label{eqn:vtr_prime}
\end{equation}
we finally obtain the ultradiscrete max-plus equations 
for the tropically discretized Sel$^{\prime}$kov model: 
\begin{eqnarray}
	X_{n+1} & = & Y_n + \max(0, 2X_n),
	\label{eqn:2-3a} \\
	Y_{n+1} & = & B - \max(0, 2X_n).
	\label{eqn:2-3b}
\end{eqnarray}

\begin{figure}[p]
\begin{center}
\includegraphics[bb=0 0 960 720, width=7.5cm]{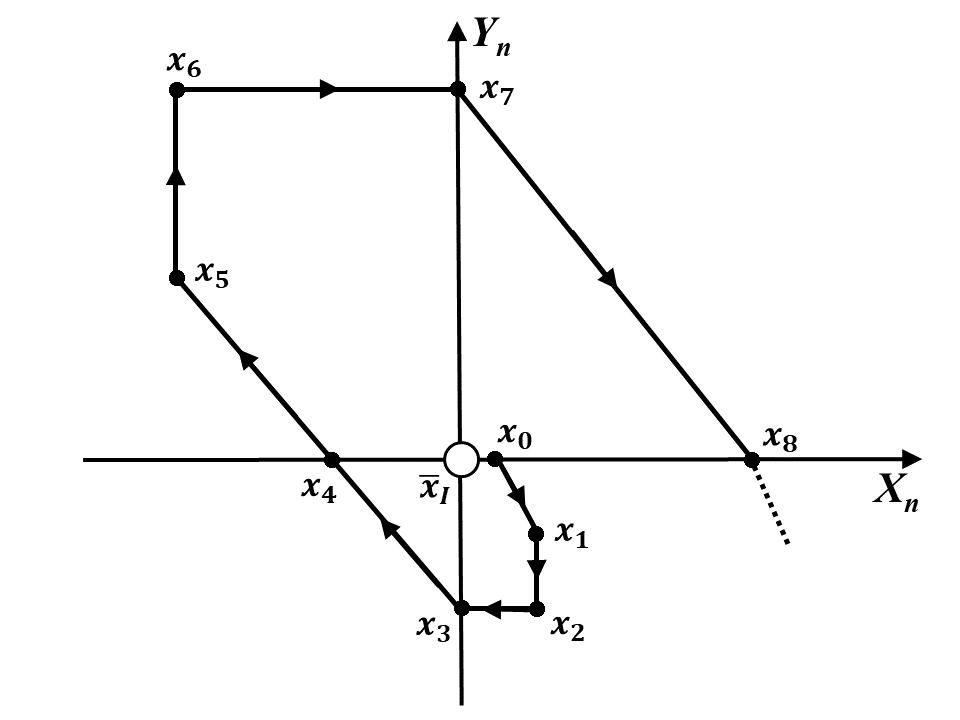}
\includegraphics[bb=0 0 960 720, width=7.5cm]{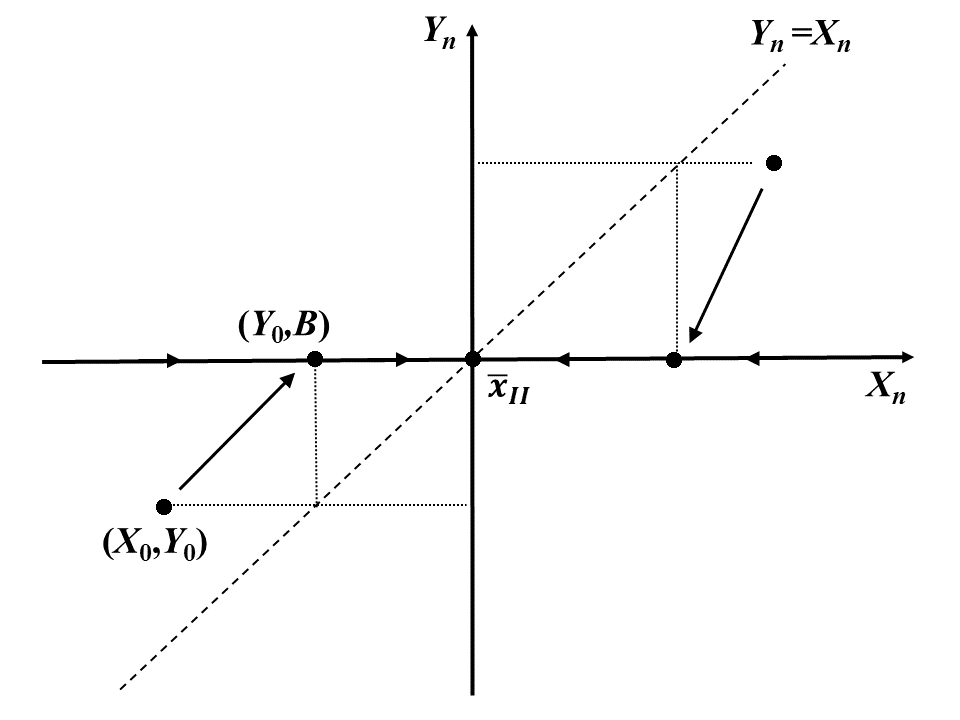}
\\
\hspace{-2cm}
(a)
\hspace{7cm}
(b)
%
\caption{\label{Fig.1} 
  (a) A trajectory in the vicinity of the unstable focus 
	$\bm{\bar x}_{I} = (B, -B)$, 
	where we set $B=0$ and $\bm{\bar x}_{0} = (1, 0)$. 
  (b) Trajectories obtained from eq.(\ref{eqn:2-2b}).}
\end{center}
\end{figure}

Now we discuss dynamical properties 
of eqs.(\ref{eqn:2-3a})-(\ref{eqn:2-3b}).
These equations are considered 
as a discrete dynamical system $\bm{x}_{n+1}=\bm {F}(\bm{x}_{n})$ 
for the state variable $\bm{x}_n=(X_n,Y_n)$ 
by equipping the evolution operator 
$ \bm {F} : (x,y) \mapsto (y+\max(0,2x), B-\max(0,2x))$.
A trajectory $\{\bm{x}_0, \bm{x}_1, \bm{x}_2, \dots \}
\left( \equiv \{ \bm{x}_n \} \right)$ 
from the initial point $\bm{x}_0 =(X_0,Y_0)$ 
is given by $\bm{x}_n = \bm {F}^n(\bm{x}_0) (n=1,2,\dots)$.
%
%
Here $(X_n, Y_n)$ plane is divided into 
the following two regions I: $X_n > 0$ 
and II: $X_n\leq 0$.
In each region, eqs.(\ref{eqn:2-3a})-(\ref{eqn:2-3b}) can be represented 
as the following matrix form.
\begin{description}
	\item[(Region I)] When $X_n > 0$, 
	eqs. (\ref{eqn:2-3a})-(\ref{eqn:2-3b}) can be rewritten as %
	\begin{eqnarray}
		\left(
   			\begin{array}{ccc}
      		X_{n+1}  \\
      		Y_{n+1}  
   			\end{array}
  		\right)
		= 
		\left(
   			\begin{array}{ccc}
      		2 & 1  \\
      		-2 & 0  \\
   			\end{array}
  		\right)
		\left(
    		\begin{array}{ccc}
      		X_{n}  \\
      		Y_{n}  
    		\end{array}
  		\right)
		+
		\left(
   			\begin{array}{ccc}
      		0  \\
      		B  
  			\end{array}
  		\right) ,
		\label{eqn:2-2a}
	\end{eqnarray} 
	where eq. (\ref{eqn:2-2a}) has the fixed point $\bm{\bar x}_I=(B,-B)$. 
	The matrix
	$
	\bm{A}_I = \left(
   			\begin{array}{ccc}
      		2 & 1  \\
      		-2 & 0  \\
   			\end{array}
  		\right)
	$
	satisfies Tr$\bm{A}_{I} = $ det$\bm{A}_I = 2$, where Tr and det stand for trace and determinant of a matrix, respectively.
	Therefore, the trajectory given by eq.(\ref{eqn:2-2a}) is characterized 
	as a clockwise spiral source\cite{Galor}, 
	whose center is the unstable fixed point $\bm{\bar x}_I$; 
	Fig. \ref{Fig.1}(a) shows a typical trajectory of $\{\bm{x}_{n}\}$. 
	
%
%

	\item[(Region II)] 
When $X_n\leq 0$, 
the matrix form of eqs. (\ref{eqn:2-3a})-(\ref{eqn:2-3b}) is 
	\begin{eqnarray}
		\left(
   			\begin{array}{ccc}
      		X_{n+1}  \\
      		Y_{n+1}  
   			\end{array}
  		\right)
		= 
		\left(
   			\begin{array}{ccc}
      		0 & 1  \\
      		0 & 0  \\
   			\end{array}
  		\right)
		\left(
    		\begin{array}{ccc}
      		X_{n}  \\
      		Y_{n}  
    		\end{array}
  		\right)
		+
		\left(
   			\begin{array}{ccc}
      		0  \\
      		B  
  			\end{array}
  		\right).
	\label{eqn:2-2b}
	\end{eqnarray} 
Equation (\ref{eqn:2-2b}) has the fixed point $\bm{\bar x}_{II}=(B,B)$.
For the matrix 
$\bm{A}_{II} = \left(
   		\begin{array}{ccc}
      0 & 1  \\
      0 & 0  \\
   		\end{array}
  	\right)$, 
	Tr$\bm{A}_{II} = $ det$\bm{A}_{II} = 0$.
The fixed point $\bm{\bar x}_{II}$ becomes a stable node.
Actually for any $\bm{x}_0=(X_0,Y_0)$, 
it is readily found that 
$\bm{x}_{1}=(Y_0,B)$ and $\bm{x}_{2}=(B,B)=\bm{\bar x}_{II}$ 
as shown in Fig. \ref{Fig.1} (b). 
\end{description}
%

\begin{figure}[h!]
	\begin{center}
		\includegraphics[height=7.5cm]{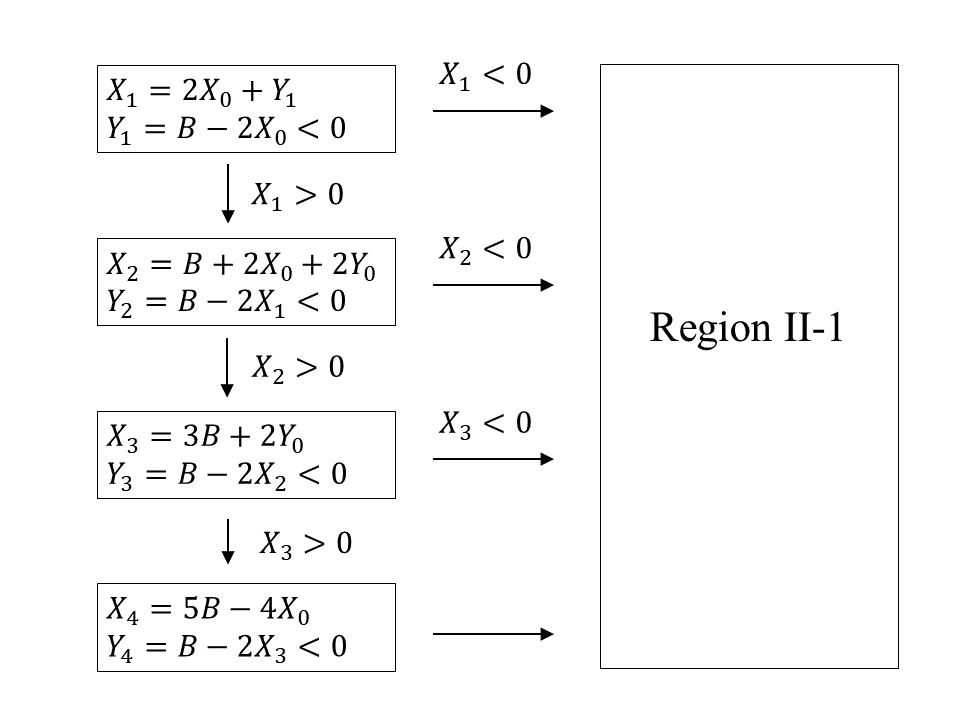}
		\caption{Flowchart for time evolutions of $\bm{x}_{n}$
		from the initial state $(X_0,Y_0)$ in region I. 
		$(X_0,Y_0)$ moves into region II-1 within four iteration steps.}
		\label{fig:flowchart}
	\end{center}
\end{figure}

\begin{figure}[h!]
\begin{center}
\includegraphics[width=7cm]{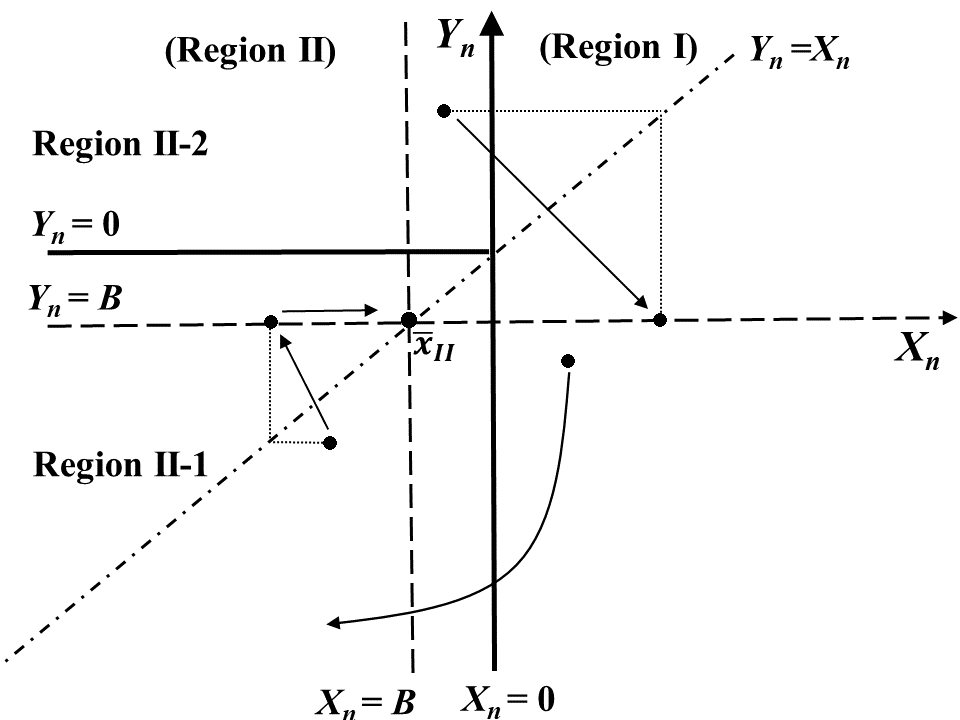}
\includegraphics[width=7cm]{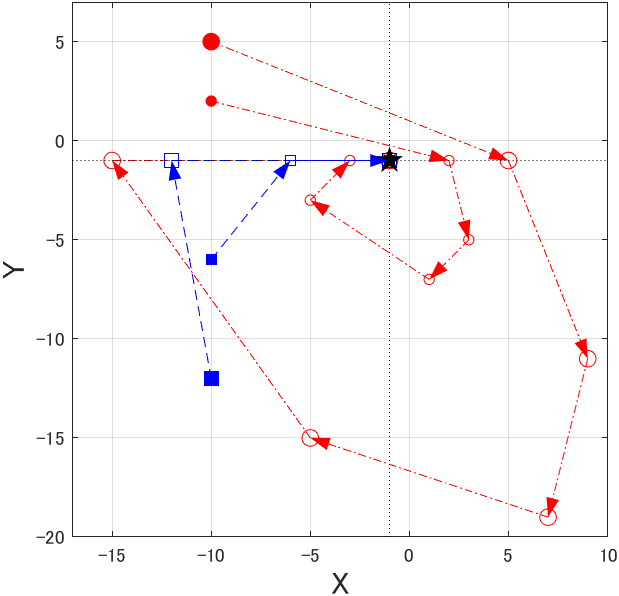}
\\
(a)
\hspace{7cm}
(b)\\
\caption{\label{Fig.2} 
	Trajectories obtained from eqs.(\ref{eqn:2-3a})-(\ref{eqn:2-3b}) 
	with $B\leq 0$.
	(a) Schematic explanation.
	(b) Numerical results from four different initial states 
	described by filled circles and squares. 
	Excitability appears when starting from the red filled circles. 
	Here we set $B=-1$.
	Trajectories proceed in the direction of the arrows 
	and finally reach the stable fixed point 
	$\bm{\bar x}_{II} = (B, B)$ shown by the black star.}
\end{center}
\end{figure}

\begin{figure}[h!]
	\begin{center}
		\includegraphics[width=7cm]{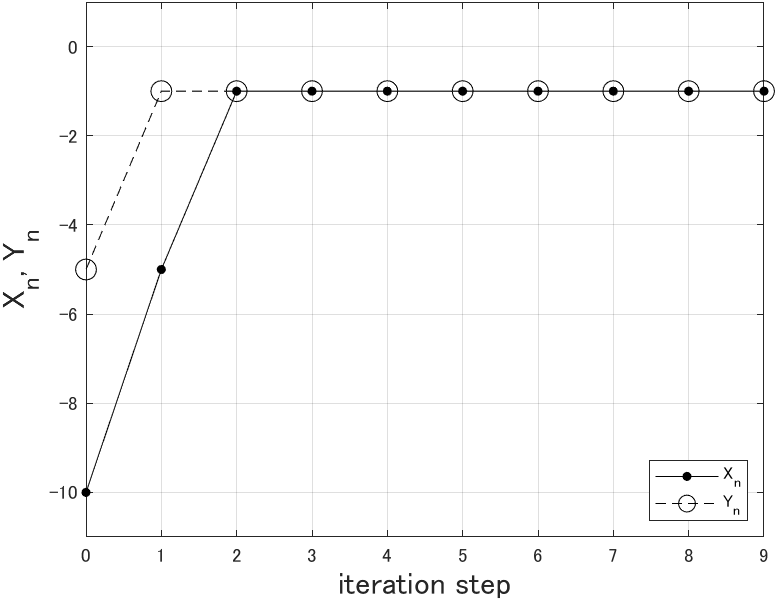}
		\includegraphics[width=7cm]{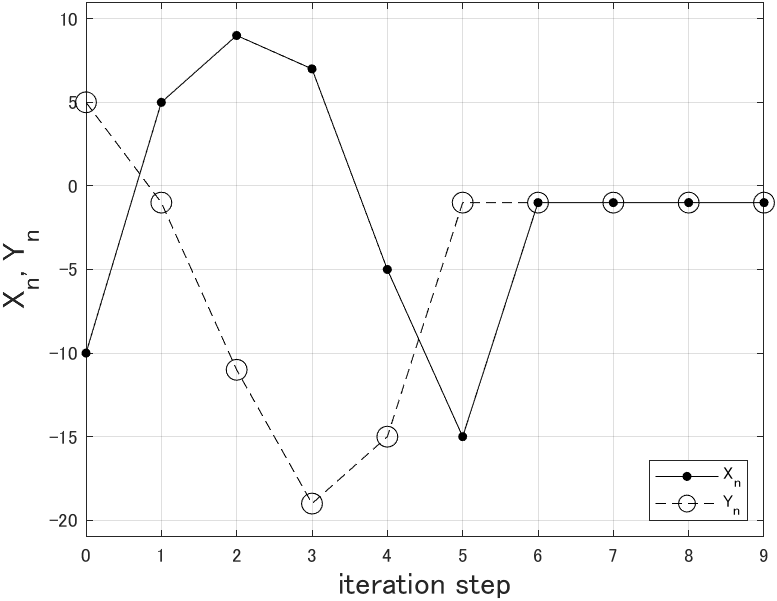}
	\\
	(a)
	\hspace{7cm}
	(b)\\
	\caption{\label{fig:excitable} 
	$(X_{n}, Y_{n})$ as a function of $n$ 
	for eqs.(\ref{eqn:2-3a})-(\ref{eqn:2-3b}) with $B = -1$ 
	from two different initial conditions.}
	\end{center}
	\end{figure}

\begin{figure}[h!]
	\begin{center}
	\includegraphics[width=7cm]{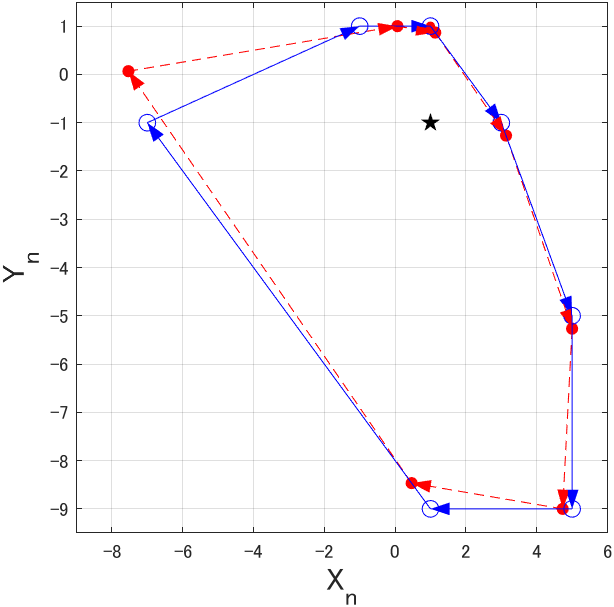}
	\includegraphics[width=7cm]{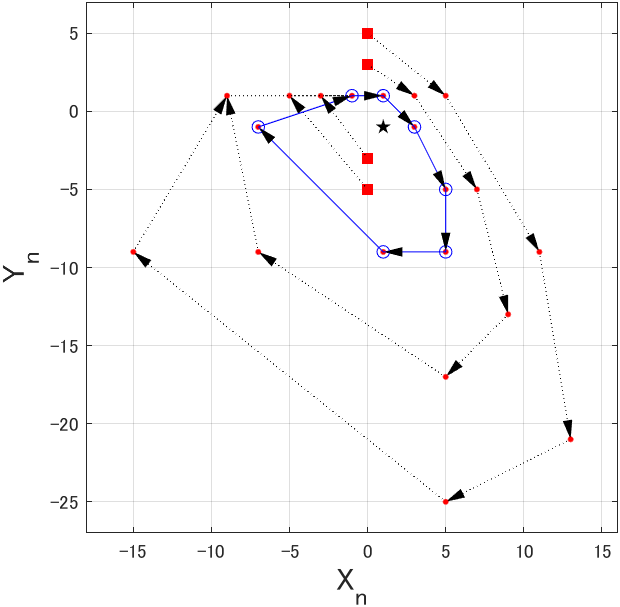}
	\\
	(a)
	\hspace{7cm}
	(b)\\
	\caption{\label{fig:lc} 
	  (a) The two limit cycles $\mathcal{C}$ (open circles) 
		and $\mathcal{C}_s$ (filled circles).
		(b) Examples of trajectories starting 
		from four different filled squares.
		The trajectories finally converge into $\mathcal{C}$ 
		consisting of the blue open circles.
		The star in each figure shows $\bm{\bar x}_{I} = (B, -B)$. 
		Here we set $B=1$.}
	\end{center}
	\end{figure}

Taking these dynamical properties in regions I and II into account, 
bifurcation of eqs. (\ref{eqn:2-3a})-(\ref{eqn:2-3b}) 
can be grasped as follows.
(i) When $B \leq 0$, both $\bm{\bar x}_{I}$ and $\bm{\bar x}_{II}$ 
are in region II; 
$\bm{\bar x}_{II}$ becomes a fixed point, but $\bm{\bar x}_{I}$ does not.
Then, eqs.(\ref{eqn:2-3a})-(\ref{eqn:2-3b}) have a unique fixed point $\bm{\bar x}_{II}=(B,B)$.
(i-a) When $\bm{x}_{0}=(X_0,Y_0)$ belongs to ``region II-1'', 
which means $X_0 \leq 0$ and $Y_0 \leq 0$, 
$\bm{x}_{1}=(Y_0,B)$ and $\bm{x}_{2}=(B,B)=\bm{\bar x}_{II}$. 
%
%
Then any initial point in region II-1 reaches 
$\bm{\bar x}_{II}$ with two iteration steps. 
(i-b) $\bm{x}_{0}$ in region I moves into region II-1
within four iteration steps as shown in Fig.\ref{fig:flowchart}.
Then, any initial point in region II-1 reaches $\bm{\bar x}_{II}$ 
within six steps. 
%
(i-c) When $\bm{x}_{0}$ is in ``region II-2'', 
$X_0 \leq 0$ and $Y_0 > 0$, 
$\bm{x}_{1}=(Y_0,B)$ belongs to region I.
Therefore, $\bm{x}_{0}$ reaches $\bm{\bar x}_{II}$ 
within seven steps from the property (i-b).
From (i-a)-(i-c), the fixed point 
$\bm{\bar x}_{II}=(B,B)$ is stable.
The trajectories of $\{ \bm{x}_{n} \}$ in this case 
are shown in Fig. \ref{Fig.2}.
Figure \ref{Fig.2} also shows that excitability appears 
when $\bm{x}_{0}$ is in region II-2.
Figure \ref{fig:excitable} shows time evolutions of 
$(X_{n}, Y_{n})$ from two different initial conditions.
It is found that excitability occurs 
in Fig.\ref{fig:excitable}(b) 
by comparing with Fig.\ref{fig:excitable}(a).
(ii) When $B>0$, $\bm{\bar x}_{I}=(B,-B)$ becomes a unique unstable fixed point.
We find that there exist two clockwise periodic solutions, 
$\mathcal{C}$ and $\mathcal{C}_s$ as shown in Fig.\ref{fig:lc}(a), 
which are composed of the following seven points, respectively: 
$\mathcal{C} = \{(B,B) \rightarrow (3B,-B) 
\rightarrow (5B,-5B) \rightarrow (5B,-9B) 
\rightarrow (B,-9B) \rightarrow (-7B,-B) 
\rightarrow (-B,B) \left[ \rightarrow (B,B) \right] \}$, 
$\mathcal{C}_s = \{ \left(\frac{B}{15},B \right) 
\rightarrow \left(\frac{17B}{15}, \frac{13B}{15} \right)
\rightarrow \left(\frac{47B}{15}, -\frac{19B}{15} \right) 
\rightarrow \left(5B, -\frac{79B}{15} \right) 
\rightarrow \left(\frac{71B}{15}, -9B \right) 
\rightarrow \left(\frac{7B}{15}, -\frac{127B}{15} \right) 
\rightarrow \left(-\frac{113B}{15}, \frac{B}{15} \right) 
\left[ \rightarrow \left(\frac{B}{15}, B \right) \right] \}$.
%
%
%
It is also found that a trajectory with any initial condition 
except for $\bm{\bar x}_{I}$ is finally absorbed 
in $\mathcal{C}$ or $\mathcal{C}_s$.
Therefore, $\mathcal{C}$ and $\mathcal{C}_s$ are limit cycles.
Figure \ref{fig:lc}(b) shows trajectories 
from four different initial conditions; 
they finally converge into $\mathcal{C}$.

Regarding basins of $\mathcal{C}$ and $\mathcal{C}_s$, 
we focus on the fact that any trajectory has a point 
with $Y_{n} = B$ at a certain iteration step $n$.
In other words, a point on 
the line $l_{B}$, $\bm{x}_n = (X_n,B)$ exists in all trajectories.
%
%
It is possible to prove that all trajectories with a point on $l_{B}$ 
except for $\left( \frac{B}{15},B \right)$ 
and $\left( \frac{17B}{15},B \right)$
are finally absorbed into $\mathcal{C}$.
And any trajectory with $\left( \frac{B}{15},B \right)$ 
or $\left( \frac{17B}{15},B \right)$ finally goes into $\mathcal{C}_s$.
Therefore, there exist only two limit cycles 
$\mathcal{C}$ and $\mathcal{C}_s$ 
in eqs.(\ref{eqn:2-3a})-(\ref{eqn:2-3b}).
%
%
%

%

\begin{figure}[h!]
	\begin{center}
		\includegraphics[height=7.5cm]{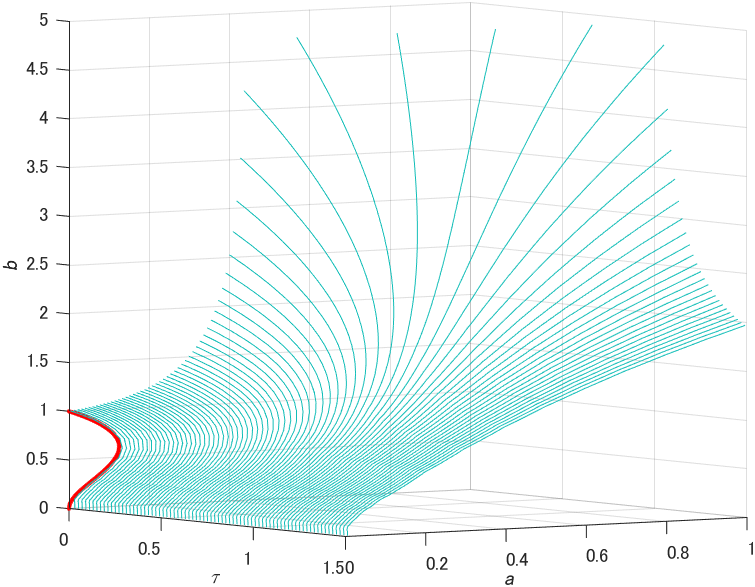}
		\caption{Neimark-Sacker bifurcation surface as a function 
		of $a$, $b$, and $\tau$. 
		The red curve shows the Hopf bifurcation curve 
		for the continuous Sel'kov model: $b = b_{\pm}(a)$.}
		\label{fig:Hopf_surface}
	\end{center}
\end{figure}

Now we focus on the linear stability for 
the tropically discretized Sel'kov model.
Equations (\ref{eqn:3-2a}) and (\ref{eqn:3-2b}) are considered 
as a discrete dynamical system 
with the three positive parameters $a$, $b$ and $\tau$.
The fixed point $(\bar{x}, \bar{y})$ for the model is 
$\displaystyle \bar{x} = b, \bar{y} = \frac{b}{b^2 + a}$, 
and the Jacobian matrix at $(\bar{x}, \bar{y})$ is
\begin{equation}
	\left(\begin{array}{cc}
		\displaystyle \frac{2\tau \bar{x} \bar{y}+1}{\tau+1} 
			& \displaystyle \frac{\tau \left(\bar{x}^2 +a\right)}{\tau+1}\\
			 & \\
		\displaystyle -\frac{2\tau \bar{x} \left(\bar{y} + b \tau\right)}{\left(\tau \left(\bar{x}^2 +a\right)+1\right)^2 } 
			& \displaystyle \frac{1}{\tau \left(\bar{x}^2 +a\right)+1}
	\end{array}\right).
	\label{eqn:Jacobian}
\end{equation}
N-S bifurcation occurs under the condition 
that the eigenvalues of this Jacobian matrix are complex 
and their absolute values are equal to 1.
This condition can be expressed by a surface 
as a function of $a$, $b$ and $\tau$; 
Fig.\ref{fig:Hopf_surface} shows the numerical results.
From this figure, the following features are confirmed.
(i) N-S bifurcation occurs for arbitrary $\tau$, 
even for $\tau \rightarrow +\infty$.
(ii) When $\tau \rightarrow 0$, N-S bifurcation curve 
is exactly consistent with Hopf bifurcation curve 
of the continuous Sel'kov model, $b = b_{\pm}(a)$. 
(iii) When $\tau \rightarrow +\infty$, 
N-S bifurcation curve becomes $b = \sqrt{a}$, 
and a limit cycle emerges when $b > \sqrt{a}$.

\begin{figure}[h!]
	\begin{center}
	\includegraphics[width=7cm]{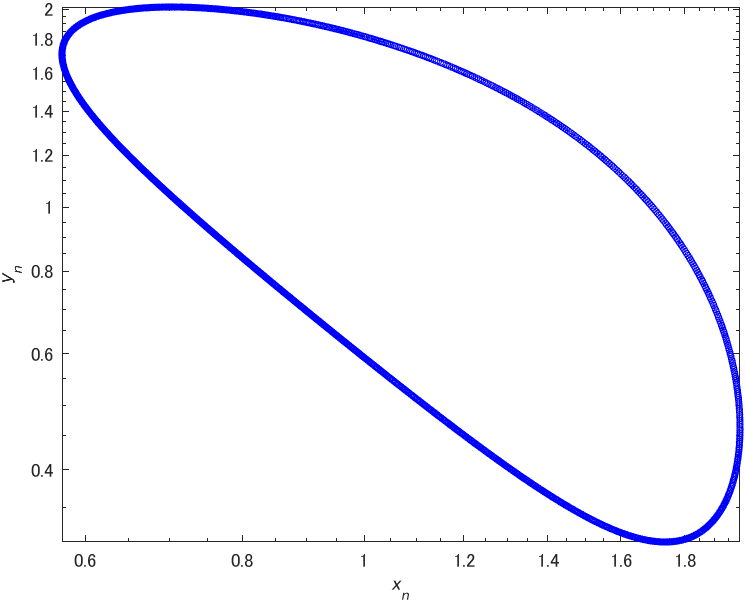}
	\includegraphics[width=7cm]{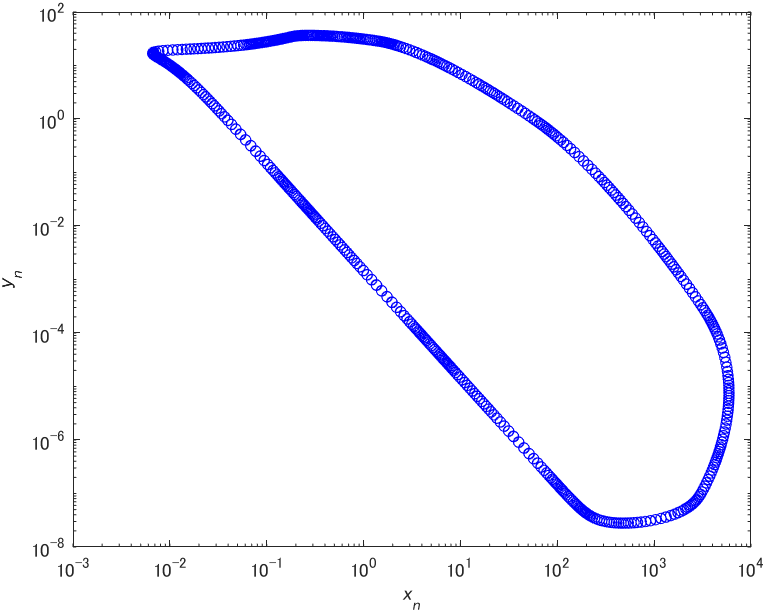}
	\\
	(a)
	\hspace{7cm}
	(b)\\
	\includegraphics[width=7cm]{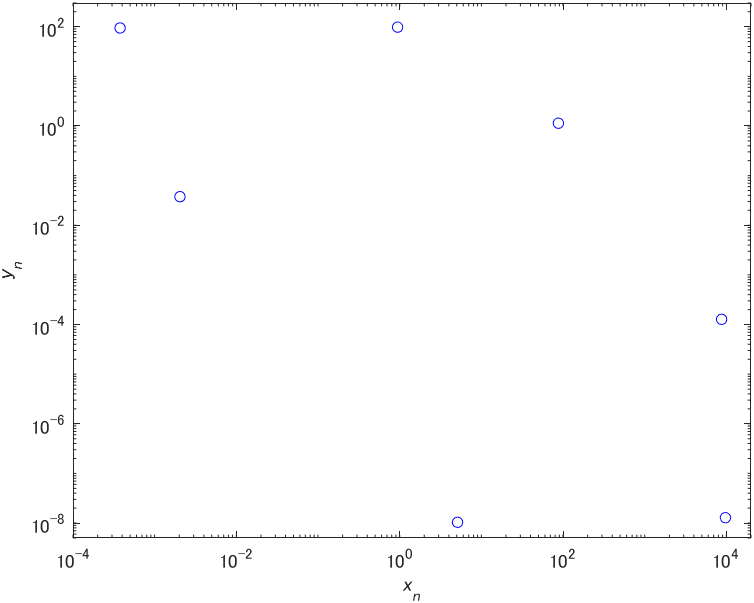}
	\includegraphics[width=7cm]{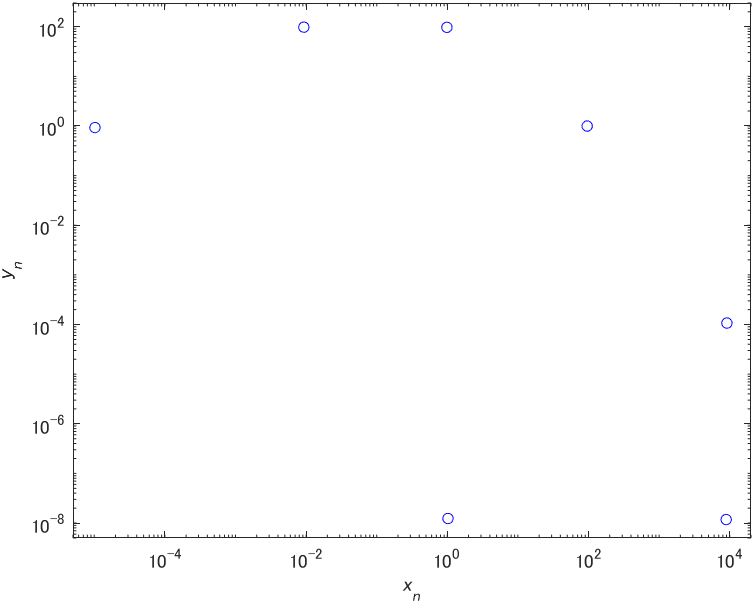}
	\\
	(c)
	\hspace{7cm}
	(d)\\
	\caption{\label{fig:trpclc} 
	  The limit cycles obtained 
		from eqs.(\ref{eqn:3-2a})-(\ref{eqn:3-2b}) 
		with $a=0.01$ and $b=0.98$.
		The values of $\tau$ are (a) 0.1, (b) 25, 
		(c) 2500, and (d) 10$^{5}$. 
		These are plotted with log-log scales.}
	\end{center}
	\end{figure}

\begin{figure}[h!]
	\begin{center}
	\includegraphics[width=7cm]{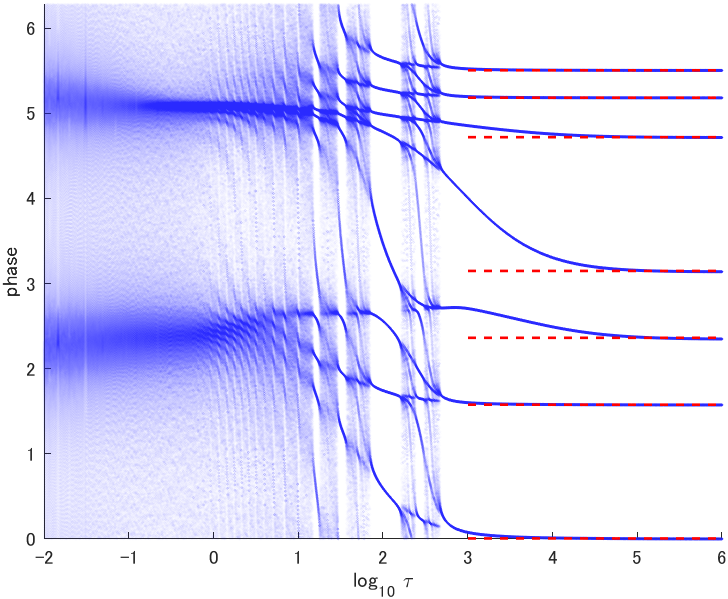}
	\includegraphics[width=7cm]{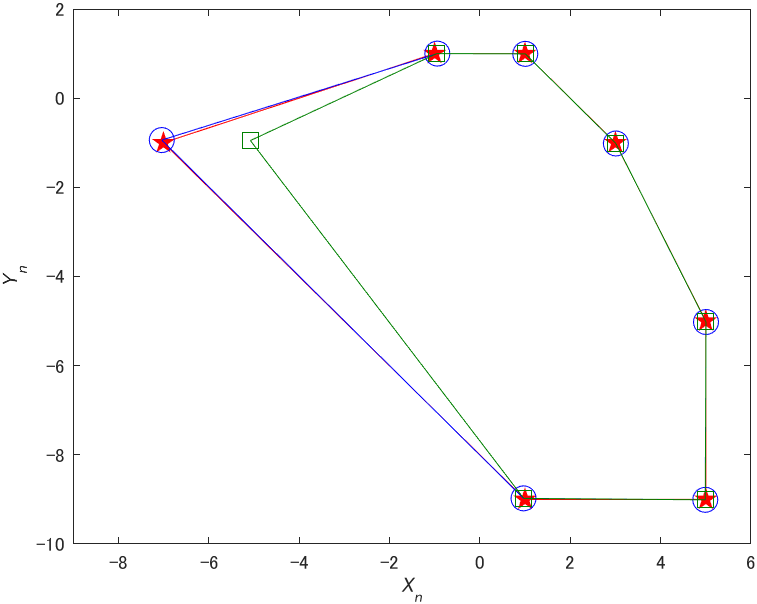}
	\\
	(a)
	\hspace{7cm}
	(b)\\
	\caption{\label{fig:lc_deltat} 
	  (a) The density of the phase $\{\theta_{n}(\tau)\}$ 
		in the limit cycles, where $a=0.01, b=0.98$. 
		$\theta_{n}(\tau)$ is defined as eq.(\ref{eqn:def_theta}). 
		The red broken line segments show $\Theta_{n}$ 
		defined as eq.(\ref{eqn:del_Ltheta}).
		(b) The seven states in the limit cycles. 
		Green open squares show $(X^{\ast}_{n}, Y^{\ast}_{n})$ 
		from eqs.(\ref{eqn:3-2a})-(\ref{eqn:3-2b}) for 
		$a=0.01, b=0.98,$ and $\tau=10^{6}$.
		Blue open circles show $(X^{\ast}_{n}, Y^{\ast}_{n})$ 
		from eqs.(\ref{eqn:3-2ia})-(\ref{eqn:3-2ib}).
		Red stars show $(X_{n}, Y_{n})$ in $\mathcal{C}$ 
		from eqs.(\ref{eqn:2-3a})-(\ref{eqn:2-3b}).}
	\end{center}
	\end{figure}

For the limit cycles, relationship between 
eqs.(\ref{eqn:2-3a})-(\ref{eqn:2-3b}) and 
eqs.(\ref{eqn:3-2a})-(\ref{eqn:3-2b}) is numerically investigated.
Here we show the case of $a=0.01$ and $b=0.98$ for example.
These values of $a$ and $b$ bring about limit cycle solutions  
in the continuous limit ($\tau \rightarrow 0$) of 
eqs.(\ref{eqn:3-2a})-(\ref{eqn:3-2b}), 
namely eqs.(\ref{eqn:3-1a})-(\ref{eqn:3-1b}).
It is found that limit cycles exist for arbitrary value of $\tau$, 
although the number of states in the cycles varies 
depending on $\tau$; 
Fig. \ref{fig:trpclc} shows four limit cycles 
with different values of $\tau$. 
The state $(x_{n}, y_{n})$ in the limit cycle is expressed 
by introducing the phase $\theta_{n}(\tau)$ as 
\begin{equation}
	\theta_{n}(\tau) = \arctan 
	\displaystyle \frac{\ln y_{n}-\ln \bar{y}}{\ln x_{n}-\ln \bar{x}}.
	\label{eqn:def_theta}
\end{equation}
$\theta_{n}(\tau)$ takes a value in the range of $[0, 2\pi]$.
Figure \ref{fig:lc_deltat}(a) shows the density 
of $\{ \theta_{n}(\tau) \}$ in the limit cycles as a function of $\tau$.
%
%
In this figure, the phase $\Theta_{n}$ of the seven states $(X_{n}, Y_{n})$
in the limit cycle $\mathcal{C}$ of the max-plus equations 
eqs.(\ref{eqn:2-3a})-(\ref{eqn:2-3b}) are also depicted 
with the red broken line segments, 
where $\Theta_{n}$ is defined as
\begin{equation}
	\Theta_{n} = \arctan 
	\displaystyle \frac{Y_{n} + B}{X_{n} - B},  
	\label{eqn:del_Ltheta}
\end{equation}
and now we set $B=1$.
Figure \ref{fig:lc_deltat}(a) shows that 
$\theta_{n}(\tau)$ seems to be in agreement with the case 
in the continuous limit (eqs.(\ref{eqn:3-1a})-(\ref{eqn:3-1b}))
when $\tau \lesssim 10^{-1}$. 
On the other hand, 
ultradiscrete feature emerges when $\tau \gtrsim 10^{3}$ 
and $\theta_{n}(\tau) \approx \Theta_{n}$ 
for $\tau \gtrsim 10^{4}$.
Actually, considering the following transformations 
\begin{equation}
	X^{\ast}_{n} = \frac{\ln x_{n} - \frac{\ln a}{2}}{\ln b - \frac{\ln a}{2}}, 
	\;\;\;
	Y^{\ast}_{n} = \frac{\ln y_{n} + \frac{\ln a}{2}}{\ln b - \frac{\ln a}{2}},
	\label{eqn:vtr_star}	
\end{equation}
$(X^{\ast}_{n}, Y^{\ast}_{n})$ is almost consistent 
with $(X_{n}, Y_{n})$ in $\mathcal{C}$ 
as shown in Fig.\ref{fig:lc_deltat}(b), where the two cases, 
$\tau = 10^{6}$ and $\tau \rightarrow +\infty$, are plotted.
From Fig.\ref{fig:lc_deltat}, 
it is concluded that the ultradiscrete max-plus equations 
(eqs.(\ref{eqn:2-3a})-(\ref{eqn:2-3b})) can be considered as 
the limiting equations of eqs.(\ref{eqn:3-2a})-(\ref{eqn:3-2b}) 
for $\tau \rightarrow +\infty$. 
Note that the limit $\tau \rightarrow +\infty$ automatically 
satisfies the condition of $T$, eq.(\ref{eqn:T-condi}).
%
%
The important point is that the dynamical structures 
of N-S bifurcation and the limit cycles retain 
for large $\tau$, even for $\tau \rightarrow +\infty$.

In the limit of $\tau \rightarrow +\infty$, 
eqs.(\ref{eqn:3-2a})-(\ref{eqn:3-2b}) become 
\begin{eqnarray}{}
	x_{n+1} = \left( a + x^2_n \right) y_n, 
	 \label{eqn:3-2ia}
\\
 y_{n+1} = \frac{b}{a+x_n^2}. 
	 \label{eqn:3-2ib}
\end{eqnarray}
After the above variable transformations 
(eqs.(\ref{eqn:3-3})and(\ref{eqn:vtr_prime})) for them, 
we obtain 
\begin{eqnarray}
	X_{n+1} & = & Y_n + \ln \left( 1 + e^{2X_n}\right), 
	\label{eqn:2-3ia} \\
	Y_{n+1} & = & B - \ln \left( 1 + e^{2X_n}\right).
	\label{eqn:2-3ib}
\end{eqnarray} 
Applying the above ultradiscretization, eq.(\ref{eqn:0}), 
to eqs.(\ref{eqn:2-3ia})-(\ref{eqn:2-3ib}), 
$\ln (1 + e^{2X}) \rightarrow \max(0, 2X)$, 
we reproduce eqs.(\ref{eqn:2-3a})-(\ref{eqn:2-3b}).
In other words, validity of eqs.(\ref{eqn:2-3a})-(\ref{eqn:2-3b}), 
or applicability of ultradiscretization, is judged whether 
the approximation $\ln (1 + e^{2X}) \approx \max(0, 2X)$ 
holds or not.

In summary, we have investigated dynamical properties 
of the max-plus equations derived 
from the tropically discretized Sel'kov model.
The max-plus equations exhibit N-S bifurcation and 
possess limit cycles.
These dynamical features are in common 
with the original discretized model.
For dynamical properties 
of the max-plus equations for N-S bifurcation and 
emergence of limit cycles, 
the present results are just for the specific model.
Further investigation will be necessary 
for generality of the present results.
We note that the derived max-plus equations can be considered 
as the limiting equations of the discrete model 
for $\tau \rightarrow +\infty$.
It is interesting that the tropical discretization gives us 
a framework for considering both the continuous limit 
($\tau \rightarrow 0$) and the ultradiscrete limit 
($\tau \rightarrow +\infty$).
Especially Figs.\ref{fig:Hopf_surface} 
and \ref{fig:lc_deltat}(a) suggest the possibility of 
discussing transition between continuous and ultradiscrete descriptions
preserving some essential dynamical structures. 
%
%
%

%

\bigskip

\noindent
{\bf Acknowledgement}

The authors are grateful to 
Prof. M. Murata, at Tokyo University of Agriculture and Technology, 
Prof. K. Matsuya at Musashino University,
Prof. D. Takahashi, 
Prof. T. Yamamoto, and Prof. Emeritus A. Kitada 
at Waseda University for useful comments and encouragements. 
%
This work was
supported by Sumitomo Foundation, Grant Number 200146.

\bigskip



\end{document}